\documentclass{aa}  

\usepackage{graphicx}
\usepackage{subfigure}
\usepackage{txfonts}
\usepackage{amsmath,amsfonts,amssymb}
\usepackage{float}
\usepackage{lscape}
\begin{document} 

\setlength{\parskip}{0pt}  

\title{
Impact of accretor size on the morphology of\\ supersonic, non-relativistic, axisymmetric\\ Bondi-Hoyle-Lyttleton accretion flows}

\titlerunning{Impact of accretor size on the morphology of supersonic Bondi-Hoyle-Lyttleton accretion flows}

\author{Shaghayegh Ashtari Jolehkaran\thanks{https://orcid.org/0009-0006-0127-6545}
  \and
  Lothar Brendel\thanks{https://orcid.org/0000-0001-7932-809X}
  \and 
  Rolf Kuiper\thanks{https://orcid.org/0000-0003-2309-8963}
}
\authorrunning{Jolehkaran et al.}

\institute{Faculty of Physics, University of Duisburg-Essen, Lotharstraße 1, D-47057 Duisburg, Germany\\
  \email{shaghayegh.ashtari-jolehkaran@uni-due.de, lothar.brendel@uni-due.de,
  rolf.kuiper@uni-due.de}
}

\date{\today}

\abstract{Fast-moving accretors are a ubiquitous phenomenon in astrophysics.
Their interaction with the surrounding gas can leave characteristic imprints on the form of morphological structures like bow shocks, Mach cones, and trails with different densities.
}{
We study how various physical processes affect the flow structure around an accretor with a one-way surface, its accretion rate, and accretion anisotropy. These processes correspond to distinct length scales: the Bondi radius, the stand-off distance of the bow shock, and the Hoyle-Lyttleton radius.
}{
We conducted adiabatic hydrodynamic simulations using a spherical coordinate grid centred on the accretor's location. By varying the accretor's (numerical) size across various scales — from much smaller than the stand-off distance to much larger than the Bondi radius — we analyse how the processes on these spatial scales affect the physics of the steady-state flow.
}{
All simulations achieve a steady state. When the accretor is smaller than the stand-off distance, a bow shock forms ahead of the object, and a nearly spherically symmetric atmosphere develops within this distance. Accretors smaller than the Hoyle-Lyttleton radius produce a Mach cone, while larger accretors exhibit a supersonic-to-subsonic flow transition on larger scales. Fully resolved simulations align with the Hoyle-Lyttleton theory, showing slightly anisotropic accretion with enhanced inflow from behind the object. In contrast, larger accretors approach the geometrical limit, with accretion primarily from the flow direction and a low-density `shadow' forming behind the object.
}{
The accretor's size greatly influences the small-scale and large-scale morphologies. 
Resolving the Hoyle-Lyttleton radius is essential for representing large-scale flow characteristics. Resolving the smaller stand-off distance is required only for studying the bow shock in front of the moving object: since the stand-off distance determines the bow shock's position, its non-resolution does not affect the larger-scale flow morphology.
}


\maketitle
%

\section{Introduction}
\label{Intro}
When a massive object moves supersonically through a gaseous medium, it can create an overdense wake through gravitational interaction with its surrounding material.
This phenomenon is observed in various scenarios, such as planets moving on eccentric orbits through a gaseous circumstellar disk, the motion of a galaxy through intracluster gas, or runaway black holes moving through the interstellar or intergalactic gaseous medium \citep{EDGAR2004843}.

Following the review by \citet{EDGAR2004843}, there are two main nomenclatures for the accretion rate ($\dot M$): the \citet{1939PCPS...35..405H} model ($\dot M_\mathrm{HL}$) describes accretion onto a gravitating point mass moving at a constant velocity through a medium of uniform density, ignoring the pressure of the gas. 
\citet{10.1093/mnras/104.5.273}, on the other hand, took the pressure into account but disregarded the accretor motion. 
In \citet{10.1093/mnras/112.2.195} an interpolation $\dot M_\mathrm{BH}$ combining the two cases was suggested.  

Over the years, various authors have numerically complemented the analytic work. Some studies employed 2D models and others 3D models. In Cartesian 2D and 3D non-axisymmetric simulations, \citet{1991A&A...248..301M} observed periodically changing shock cone shapes. However, in 3D simulations, the shape of the shock cone remains more stable \citep{1994ApJ...427..351R}.

Other investigations have focused on non-axisymmetric 3D simulations, in which angular momentum is directly impacted. In contrast, the \citet{10.1093/mnras/104.5.273} accretion model assumes a pressure-less medium and nullifies the gas particles' tangential momentum upon reaching the downstream axis. This construction results in the accreted matter having zero angular momentum. The density distribution and shock cone shape were influenced by the velocity gradient, leading to higher density maxima along the cone's side with lower velocities. 

\citet{1994ApJ...427..342R, 1994A&AS..106..505R, 1997A&A...317..793R, 1999A&A...346..861R} conducted significant work on \citet{10.1093/mnras/104.5.273} accretion in the Newtonian case. Across a series of papers, they systematically investigated the influence of various factors on flow properties. These factors included the accretion radius, the adiabatic index of the ideal gas, and perturbations in both density and velocity upstream of the accretor.

In an early numerical paper \citep{1994ApJ...427..351R}, the 3D Bondi-Hoyle accretion of an ideal gas with $\gamma = 5/3$ around a sphere moving at Mach 3 was investigated. 
In a subsequent paper, \cite{1994A&AS..106..505R}, investigated Bondi-Hoyle accretion for three different Mach numbers (0.6, 1.4, and 10). He found that the flow became dynamically unstable for high Mach numbers and small accretor sizes, while lower Mach numbers with the same accretor size produced stable flows. For higher Mach numbers with larger accretor sizes, the flow was also stable. His paper introduced a correction to $\dot{M}_{\mathrm{BH}}$, introducing the parameter $\beta$, which depends on the adiabatic index, $\gamma$ (see Eq.~12 in their paper).



Unlike Ruffert, some studies delved into relativistic scenarios \citep{1998ApJ...494..297F, 1998MNRAS.298..835F}. 
In these scenarios, the gas has velocities near the speed of light at infinity, requiring the incorporation of relativistic effects. 
Additional studies by various authors incorporated complex physical processes into \citet{10.1093/mnras/104.5.273} accretion simulations. These processes included radiative heating \citep{1990ApJ...356..591B} and radiation pressure \citep{1991ApJ...371..696T}. While these additions introduced complexity to the models, they were aimed at capturing a more realistic representation of the astrophysical conditions, even though they could potentially complicate the interpretation of the results.

We conducted hydrodynamic simulations of a gravitating body with a one-way surface moving supersonically through a homogeneous medium.
We focused on the axisymmetric accretion rate resulting from gravitational interaction between the medium and the body using direct numerical modelling. The visualization demonstrates that when a massive object moves through an initially stationary medium, it creates a bow shock and a Mach cone with a higher density than the undisturbed medium on the upstream side. The resulting gravitational force exerted by the medium from behind the object slows the object's movement \citep{1943ApJ....98...54C,2016A&A...589A..10T} and is known as dynamical friction. In the present work, we derived the stationary state flow around the object for a given velocity and did not follow the temporal evolution due to dynamical friction.

We varied an important physical parameter of the problem, namely the Mach number of the flow, as well as a numerical parameter, namely the radius of the inner sink cell.
Furthermore, by performing a simulation with a higher accretor mass, we confirm the analytically expected scaling of the flow properties with the Hoyle-Lyttleton radius.
It is important to note that sink cells represent, but do not accurately depict, the real astrophysical object sizes in these simulations. This means that the true size of the astrophysical object is not resolved in the simulations. Sink cells are used to represent gravitating objects without resolving their detailed structure. Nevertheless, we denote the sink cell radius as the accretor size in the following.  
We determined the consequences of not resolving particular spatial scales on the resulting larger-scale morphology, that is, the impact of the choice of the inner radius ($R_\mathrm{min}$) of the spherical coordinate system.

The physics of the accreting system under investigation can be categorized by the length scales associated with the different physical processes. 
Most of the following length scales stem from comparing the typical velocity of the different physical processes with the escape speed,
\begin{equation}
    v_\mathrm{esc} = \sqrt{2 G M / R}
,\end{equation} 
of the gravitational object.
In decreasing order, these different velocities are the speed of light ($c$), the relative velocity ($v_{\infty}$)between ambient gas and the object, and the speed of sound ($c_\mathrm{s}$). 
The associated length scales in increasing order are the Schwarzschild radius ($R_\mathrm{Sch}$), the Hoyle-Lyttleton radius ($R_\mathrm{HL}$), and the Bondi radius ($R_\mathrm{Bondi}$):
\begin{align}
    R_\mathrm{Sch}   &= \frac{2GM}{c^2} \\
    \label{eq:defRdyn}
    R_\mathrm{HL}   &= \frac{2GM}{v_{\infty}^2} \\
    R_\mathrm{Bondi} &= \frac{2GM}{c_\mathrm{s}^2}.
\end{align}
The Hoyle-Lyttleton radius, which is sometimes just called the accretion radius, gives an estimate of the critical impact parameter of the incoming gas and acts as a dividing line between accreting gas ($R < R_\mathrm{HL}$) and escaping gas ($R > R_\mathrm{HL})$. 
As supersonic, non-relativistic flows, all the cases presented here by definition fulfil the relation $R_\mathrm{Sch} \ll R_\mathrm{HL} < R_\mathrm{Bondi}$.
Another relevant length scale is the stand-off distance ($R_\mathrm{so}$), which is the position of the bow shock on the upstream axis.
In all cases presented here, the stand-off distance is smaller than the Hoyle-Lyttleton radius, meaning $R_\mathrm{Sch} \ll R_\mathrm{so} < R_\mathrm{HL} < R_\mathrm{Bondi}$.

In a parameter study, we varied the size of the accretor (actually the sink cell, see above) from scales much smaller than the stand-off distance to scales much larger than the Bondi radius. 
When choosing a size larger than one of the length scales summarized above, we can assume that the associated physical process is no longer taken into account appropriately. 
So by varying the size of the accretor across the multitudes of length scales, we are able to determine which of these processes shape the flow morphology on larger scales. 
As an example, when the accretor is much larger than the Bondi radius, we can expect the final accretion rate to be given by the so-called geometric accretion rate, because none of the small-scale physical processes are resolved on the computational domain any more, and the accretor simply captures the gas mass by its geometrical extent without any gravitational effect.
To save computational resources, we have not investigated the physics of the smallest scales, those close to the Schwarzschild radius. Consequently, all simulations stay firmly in the non-general-relativistic regime.


When varying the size of the accretor across several length scales, two well-known analytical mass accretion rates are relevant for the flow morphology on larger scales.\ The first is the Hoyle-Lyttleton accretion rate.\ It depends on $R_{\mathrm{HL}}$, which is the distance from the centre at which the flow speed is sufficient to escape the body's gravitational field. This scenario is given by \citet{1939PCPS...35..405H} and is expressed as

  \begin{equation}
     \dot{M}_{\mathrm{HL}} = \rho_{\infty} v_{\infty} \pi R_\mathrm{HL}^{2}\,.
     \label{HL-accretion}
   \end{equation}It disregards gas pressure and considers purely Keplerian trajectories of the gas particles, up to their `collision' on the negative $z$-axis, when they lose their tangential momentum and the corresponding energy.

The second is the geometric accretion rate ($\dot{M}_{\mathrm{geo}}$). It is based on the geometrical size of the object, which is effectively $R_{\mathrm{min}}$ in our case:
  
\begin{equation}
     \dot{M}_{\mathrm{geo}} = \rho_{\infty} v_{\infty} \pi R_{\mathrm{min}}^{2} \,.
     \label{geo-accretion}
\end{equation}This picture disregards gas pressure and gravity and therefore treats the trajectories of gas particles as straight lines in the accretor's rest frame.

The paper is organized as follows: In Sect.~\ref{Methods} we briefly present the system of equations of hydrodynamics written as a hyperbolic system of conservation laws and the numerical code, along with the initial setup of the simulations. In Sect.~\ref{results} we present the simulations of gravitating objects for various physical parameters and discuss the results of the simulations. Finally, in Sect.~\ref{summary} we summarize the main conclusions of this work. 

\section{Methods} \label{Methods}

\subsection{Scenario and setup}
The accretor has a known mass, $M$, moving supersonically but non-relativistic with a velocity $v_{\infty}$ through a uniform gaseous background. This ambient medium is characterized by the following properties: gas mass density ($\rho_{\infty}$), thermal pressure ($p_{\infty}$), sound speed ($c_{\infty}$), and temperature ($T_{\infty}$), all of which are measured far from the object. 

Here, the gas was assumed to behave as an ideal gas and adiabatically with an index of $\gamma= 5/3$.
We neglected any viscosity of the gas and solely considered the gravitational force exerted by the accretor, which is no self-gravity of the gas. 
We did not account for radiative heating and cooling or the influence of magnetic fields.

We treated the object as a body that permits accretion through its `surface'.
The minimum radius of the associated sink cell is in all simulations larger than the Schwarzschild radius. By decreasing this radius in a simulation series, we checked for numerical convergence of the results (i.e. even smaller $R_\mathrm{min}$ would not change the results).
Furthermore, determining the largest $R_\mathrm{min}$, which still guarantees converged results is one of the main aims of this work.

After the transition from the initial conditions to the steady state, we determined the final accretion rate, analysed the density distribution, and visualized the flow geometry. 
We investigated the formation of a bow shock resulting from the motion of the accretor through the ambient medium. 
Under the approximations described above, the flow morphology is independent of the value of the uniform density, and its temperature and pressure are taken into account by the Mach number.
Regarding it as an interstellar medium, its parameters are chosen as $\rho_{\infty} = 1.67 \times 10^{-25}~\mathrm{g}/\mathrm{cm}^3$, $p_{\infty} = 1.6 \times 10^{-13}$ erg/s, and $T_{\infty} = 6000~ \mathrm{K}$. 
The mass of the gravitating accretor is set to $M = 1 \mbox{ M}_\odot\ $, but the flow morphology can be scaled to any arbitrary mass (see Appendix~\ref{sect:mass-scaling} for details).
Our analysis involves calculating the accretion rate for different $R_{\mathrm{min}}$ (see Sect.~\ref{Accretion rate}), exploring the density distribution, examining the velocity field, and studying the accretion anisotropy for two scenarios, with Mach numbers $\mathcal{M}_\infty = 10$ and $1.5$ (Sect.~\ref{Result10} and Sect.~\ref{Result1.5}). Table~\ref{Parameter} gives an overview over the variations of $R_{\mathrm{min}}$.

\subsection{Simulation and numerics}
The hydrodynamic equations to be solved are
   \begin{equation}
   \label{eq:cont}
      \frac{\partial \rho }{\partial t} + \nabla \cdot (\rho \boldsymbol{\varv}) = 0 \,,
   \end{equation}
   \begin{equation}
      \left(\frac{\partial }{\partial t} + \boldsymbol{\varv} \cdot \nabla \right) \boldsymbol{\varv} =  - \frac{1}{\rho} \nabla p - \nabla \phi \,,
   \end{equation}
   \begin{equation}
      \frac{\partial E }{\partial t} + \nabla \cdot [(E + p) \boldsymbol{\varv}] = - \rho \boldsymbol{\varv} \cdot \nabla \phi \,,
      \label{CL2}
   \end{equation}
    \begin{equation}
      \frac{p}{\rho^\gamma} = \text{const.} \,,
      \label{adiabatic}
   \end{equation}
where $\rho$ is the mass density, $\boldsymbol{\varv}$ the velocity, $p$ the pressure of the gas,  and $\phi = - G M / r$ is the gravitational potential. Equation~\eqref{adiabatic} describes the behaviour of the gas during an adiabatic process, with $\gamma = 5/3$ being the constant adiabatic index.
The energy density ($E$) of the gas, which contains the specific thermal energy ($e_\text{th}$) and the specific kinetic energy, is
    \begin{equation}
      E = \rho \left(e_\text{th} + \boldsymbol{\varv}^2/2\right)\,.
    \end{equation}
The specific thermal energy ($e_\text{th}$) in turn is related to the pressure:
   \begin{equation}
     p = (\gamma - 1)\rho e_\text{th}  \,.
   \end{equation}
Given the initial values of density ($\rho_{\infty}$) and pressure ($p_{\infty}$) for the homogeneous medium, the adiabatic index ($\gamma$), and the speed ($v_{\infty}$) of the moving object, the adiabatic speed of sound ($c_{\infty}$) is given as
   \begin{equation}
     c_{\infty} =  \sqrt{\gamma \frac{p_\infty}{\rho_\infty}} \,,
   \end{equation}
which in turn determines the Mach number of the system:
   \begin{equation}
     \mathcal{M} =  \frac{v_{\infty}}{c_{\infty}} \,.
   \end{equation}

We used the open-source software PLUTO, version 4.4 \citep{Mignone} to solve the Eqs.~\eqref{eq:cont}-\eqref{adiabatic}. PLUTO is a versatile tool that allows for Riemann problems to be resolved at cell interfaces and therefore numerical fluxes to be calculated. It is recognized for employing `high-resolution shock-capturing' methods. In our numerical simulations, we chose a second-order Runge-Kutta time-stepping scheme for time integration. The spatial discretization involves a linear reconstruction method, where the values within each computational cell are estimated using linear interpolation based on the known values at the cell vertices. For solving the Riemann problem, we utilized the Harten-Lax-van Leer (HLL) approximate Riemann solver. The HLL solver efficiently computes the fluxes at the cell interfaces, providing a robust and accurate treatment of fluid dynamics, especially in the presence of shock waves or other discontinuities.

As initial conditions, we chose an axisymmetric flow in the negative $z$-direction. We employed a 2D spherical grid $(r, \theta)$ to represent our simulation domain. The polar direction extends from $0$ to $\pi$ with uniform angular spacing, where $0$ represents the positive $z$-direction and $\pi$ the negative $z$-direction. The radial direction is limited from below by $R_{\mathrm{min}}$ (see Table \ref{Parameter} for its different values). The outer radius of the computation domain ($R_{\mathrm{max}}$) was set to $20$ pc. In terms of boundary conditions, for the upper hemisphere $\theta\in[0,\pi/2]$ we enforced a mass-inflow of density ($\rho_{\infty}$) and velocity ($v_{\infty}$) in the negative $z$-direction. For the lower hemisphere $\theta\in[\pi/2,\pi]$, a zero-gradient condition was applied to enable outflowing material to exit the computational domain without reflections. 
The boundary condition at $R_\mathrm{min}$ was set to a one-way condition, meaning the gas was allowed to freely enter the central sink cell (zero gradient boundary condition), but no flow was allowed from the sink into the computational domain (reflective boundary condition). 
The simulations with Mach 1.5 used a default grid configuration of $N_r \times N_\theta = 682 \times 100$ cells, which means the number of grid cells was kept constant.
In the simulation series with Mach 10, we changed the number of grid cells in the radial direction according to the minimum radius of the computational domain to keep the spatial resolution of the different simulations constant in regions present in all simulations.

The simulations can therefore be considered as the accretor being stationary at the coordinate system's origin. To achieve high spatial resolution around the body surface and the region of interest, we used a logarithmic grid spacing in the radial direction, with increasing grid cell lengths towards larger scales. When varying $R_{\mathrm{min}}$, we chose different values for the number of grid cells in order to keep the grid structure identical on the included scales.

\begin{table}
\label{Parameter}
\caption{Simulation overview.}
\begin{tabular}{clllc} \\
\hline \hline
\noalign{\smallskip}
$\mathcal{M}_\infty$ & $R_{\mathrm{min}}$ [pc] &  $R_{\mathrm{min}} / R_{\mathrm{HL}} $ & $N_r$ & Figures   \\
\noalign{\smallskip} \\
\hline 
\noalign{\smallskip}
1.5 & $5 \times 10^{-7}$ &  $2.32 \times 10^{-2} $ & 682 & Fig.~\ref{fig:accretion}\\
    & $10^{-6}$ &  $4.64 \times 10^{-2} $ &  & Figs.~\ref{Analysis}, \ref{fig:accretion} \\
    & $3 \times 10^{-6}$ & $1.39 \times 10^{-1} $ & & Figs.~\ref{Analysis}, \ref{fig:accretion}\\
    & $8 \times 10^{-6}$ & $3.71 \times 10^{-1} $ & & Fig.~\ref{fig:accretion}\\
    & $1.5 \times 10^{-5}$ & $6.96 \times 10^{-1}$ & & Fig.~\ref{fig:accretion}\\
    & $1.7 \times 10^{-5}$ & $7.89 \times 10^{-1}$ & & Fig.~\ref{fig:accretion}\\
    & $1.95 \times 10^{-5}$ & $9.05 \times 10^{-1}$ & & Figs.~\ref{Analysis}, \ref{fig:accretion}\\
    & $5.7 \times 10^{-5}$ & $2.64$  & & Figs.~\ref{Analysis}, \ref{fig:accretion}\\
    & $10^{-4}$ & $4.64$ & & Figs.~\ref{Analysis}, \ref{fig:accretion}\\
    & $10^{-3}$ & $4.64 \times 10$ & & Figs.~\ref{Analysis}, \ref{fig:accretion}\\
    & $10^{-2}$ & $4.64 \times 10^{2}$ & & Figs.~\ref{Analysis}, \ref{fig:accretion}\\
\noalign{\smallskip} \\
\hline 
\noalign{\smallskip}
 10 & $5 \times 10^{-9}$ & $ 9.65 \times 10^{-3} $ & 892 & Fig.~\ref{fig:accretion}\\
    & $2 \times 10^{-8}$ & $3.86 \times 10^{-2} $ & 682 & Figs.~\ref{morphology}(a),  \ref{Analysis}, \ref{fig:accretion}\\
    & $7 \times 10^{-8}$ & $1.35 \times 10^{-1}$ & 610 & Figs.~\ref{morphology}(b), \ref{Analysis}, \ref{fig:accretion}\\
    & $2.6 \times 10^{-7}$ & $5.04 \times 10^{-1}$ & 582 & Figs.~\ref{morphology}(c), \ref{Analysis}, \ref{fig:accretion}\\
    & $10^{-6}$ & $1.93$ & 555 & Figs.~\ref{morphology}(d), \ref{Analysis}, \ref{fig:accretion}\\
    & $10^{-5}$ & $1.93 \times 10$ & 460 &  Figs.~\ref{morphology}(e), \ref{Analysis}, \ref{fig:accretion},\\
    & $10^{-4}$ & $1.93 \times 10^{2}$ & 390 & Figs.~\ref{morphology}(f),  \ref{fig:accretion}\\
    & $10^{-3}$ & $1.93 \times 10^{3}$ & 320 & Figs.~\ref{Analysis}, \ref{fig:accretion} \\
\noalign{\smallskip} \\
\hline
\end{tabular}
\tablefoot{
Each line represents an individual simulation with the corresponding parameters. Columns include the Mach number, the inner radius of the computational domain, and the radius normalized by the Hoyle-Lyttleton radius, where $R_\mathrm{HL} = 2.155 \times 10^{-5}$ pc for Mach 1.5 and $R_\mathrm{HL} = 5.177 \times 10^{-7}$ pc for Mach 10. Thus, $R_\mathrm{HL}$ is several orders of magnitude smaller than $R_\mathrm{max} = 20$ pc, making the simulation unaffected by numerical artefacts related to the finite domain size.
The second to last column gives the number of grid cells in the radial direction $N_r$, and the number of grid cells in the polar direction $N_\theta = 100$ is the same for all simulations performed.
The last column indicates the figures, in which the simulation data have been used. The final column shows the figures in which the simulation data are presented.
}
\end{table}



\section{Results} \label{results}

\subsection{Flow morphologies} 
\subsubsection{Results for Mach number 10}\label{Result10}
\begin{figure*}[htbp]  \centering\includegraphics[width=1\textwidth]{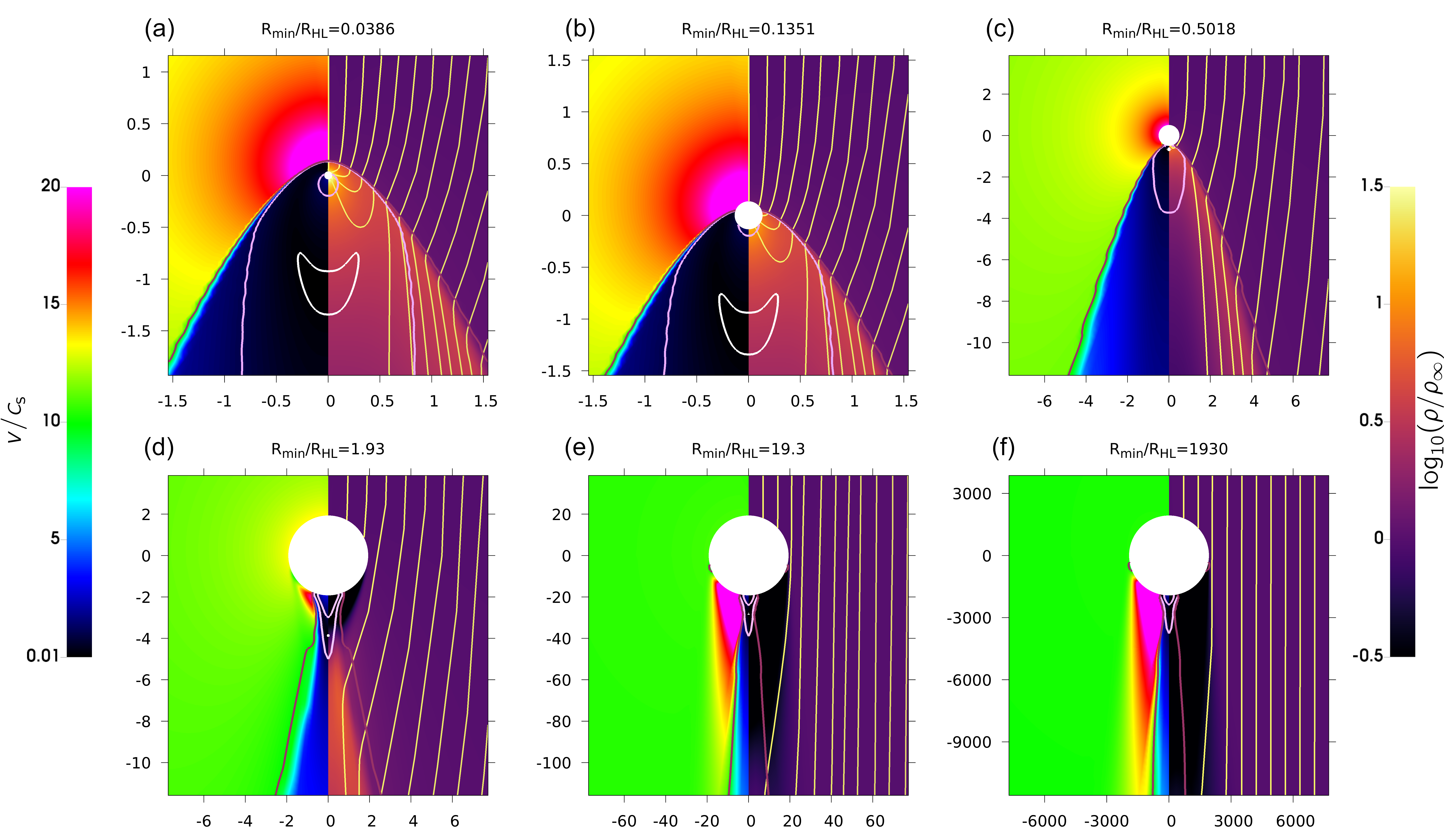}
    \caption{Fields of velocity ($v$) normalized by the local sound speed ($c_\mathrm{s}$; left half of the panels) and density (right half) for different $R_\mathrm{min}$ and a Mach number $\mathcal{M}_\infty=10$.\ The units of both axes are $R_\mathrm{HL}$. Overlaid in yellow on the density are selected gas trajectories.  Three contours of constant $v/c_\mathrm{s}\in\{0.1,1,10\}$ are shown in white, pink, and purple, respectively.}
    \label{morphology}
\end{figure*}

Figure \ref{morphology} illustrates the steady-state solutions for the density and the velocity field in simulations with a Mach number of 10 for different sizes $R_{\mathrm{min}}$ of the accretor.
The basic features of these flow morphologies as a function of the accretor size as well as the change in behaviour of the bow shock with increasing $R_\mathrm{min}$ are in agreement with the earlier work by \citet{1994A&AS..106..505R}.

The gas moves with the velocity $v_{\infty}$ at large upstream distances towards the accretor, inducing significant density changes in its vicinity. A bow shock forms -- in front of the object for the simulation with the smallest $R_{\mathrm{min}}$, but behind the object for the largest $R_{\mathrm{min}}$ -- decelerating material from supersonic to subsonic speeds and consequently causing a density increase.


Figure~\ref{Analysis} presents the the relative density (panels a and b) and velocity (c and d) along the $z$-axis. They allow for a complimentary analysis.
In the upstream direction, $z>0$, we observe an increase in the density compared to the inflowing gas density. But only for the smallest $R_\mathrm{min}$ it includes the sudden density increase at the shock front, the position of which defines the stand-off distance $R_\mathrm{so}$. 
The data for larger $R_\mathrm{min}$ are consistent with the resolved case in the upstream direction, they only miss the corresponding bow shock region. From Fig.~\ref{morphology} we already know that the bow shock does actually not disappear, but its front location is shifted towards smaller $z$-values: For $R_\mathrm{min}=0.1351R_\mathrm{HL}$, the shock front intersects the accretor, for still larger $R_\mathrm{min}$, it is even located behind.

From the incoming flow towards the object and further downstream, we observe the following: For the smallest $R_\mathrm{min}$, the Mach number starts at the ambient value of the incoming flow, peaks at the shock front, and then gradually decreases as it reaches the object. For $R_\mathrm{min}$ greater than $R_\mathrm{HL}$, there is no significant change in the Mach number upstream because the bow shock is absent in front of the object. Downstream, the Mach number increases due to gravitational acceleration and then eventually returns to the ambient Mach number at a significant distance from the object.

Figure~\ref{Analysis}(e) shows an anisotropic accretion analysis. Anisotropic accretion refers to the non-uniform or direction-dependent accretion of material onto an accretor.
The case of spherical symmetric accretion would yield a constant value of unity independent of the angle. In the other extreme case of geometric accretion, the maximum value of 4 corresponds to the ratio of the accreting spherical surface area $(4 \pi R_\mathrm{min}^2)$ to the accretors projected area $(\pi R_\mathrm{min}^2)$. 
For the resolved case of $R_\mathrm{min}$ smaller than $R_\mathrm{HL}$, the simulation resembles the theoretically expected case in which a slightly higher accretion is coming from behind the object than from the upstream direction. 
As seen in comparison with the $\mathcal{M}_\infty=1.5$ case (Sect.~\ref{Result1.5}), the level of anisotropy increases with a higher Mach number of the flow. This is expected, because, for higher Mach numbers, the kinetic effects become more dominant than the pressure effects, which effectively yields a behaviour closer to Hoyle-Lyttleton than Bondi-Hoyle accretion.
But in the case of $R_\mathrm{min}$ greater than $R_\mathrm{HL}$, the accretion becomes not only more anisotropic but importantly in the opposite direction: Now the majority of accretion comes from the upstream direction. This behaviour is actually expected in the geometrical limit.
This panel also demonstrates that by decreasing the radius in a simulation series, numerical convergence of the results is achieved (i.e. further reducing $R_\mathrm{min}$ does not alter the outcomes). Additionally, identifying the largest $R_\mathrm{min}$ that still ensures convergence is a key objective. For instance, in the case of Mach 10, the two smallest accretor sizes do not exhibit different accretion flows. As shown in Fig.~\ref{Analysis}(a)-(d), the purple and dark blue lines overlap exactly, with the only difference being the size of the object ($R_\mathrm{min}$), that is, the purple lines in those panels are getting deeper towards the accretor, which is also seen by the different values of the accretion rates for those two cases in panel (e), measured at their specific $R_\mathrm{min}$.


\subsubsection{Results for Mach number 1.5}
\label{Result1.5}
\begin{figure*}[htbp]
        \centering
        \begin{minipage}[t]{0.45\textwidth}
                \centering
        \textbf{$\mathcal{M}_\infty = 10$}
                \includegraphics[width=0.85\textwidth]{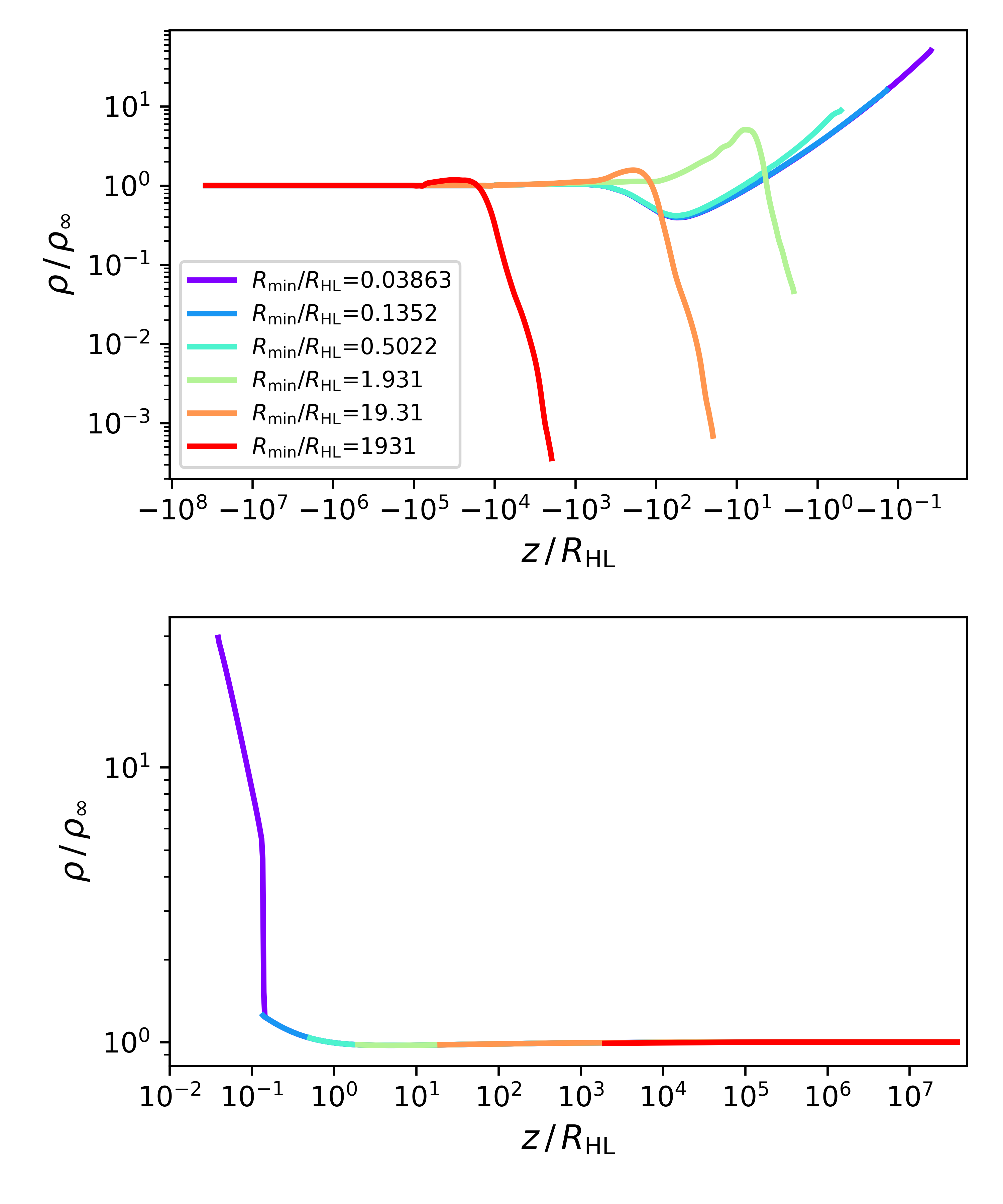}
                \parbox[b]{0.05em}{(a) \\[24ex] (b) \\[3ex]} \\
                \vspace{0.1cm}
                \includegraphics[width=0.85\textwidth]{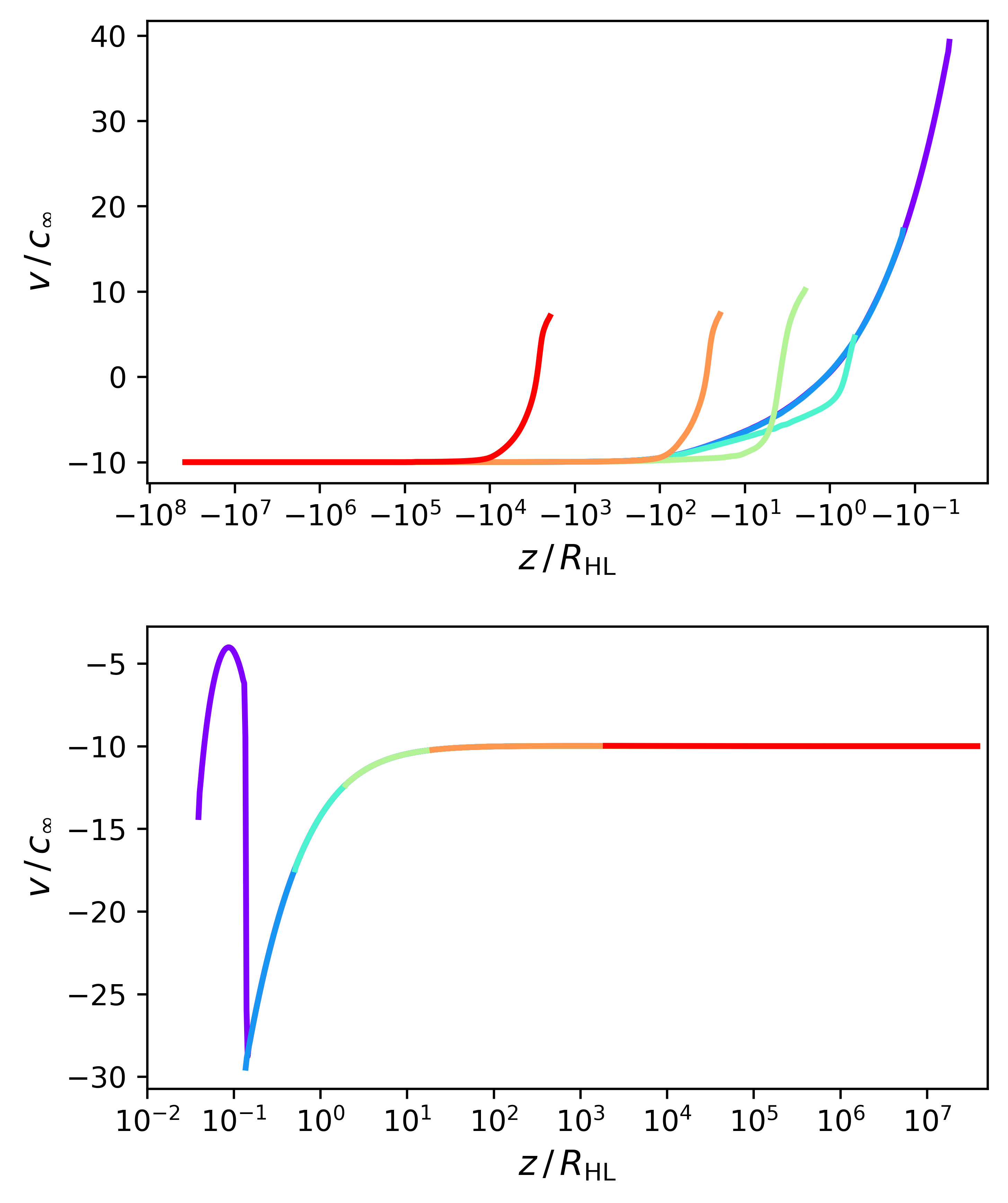}
                \parbox[b]{0.05em}{(c) \\[25ex] (d) \\[2ex]} \\ 
                \vspace{0.1cm}
                \includegraphics[width=0.85\textwidth]{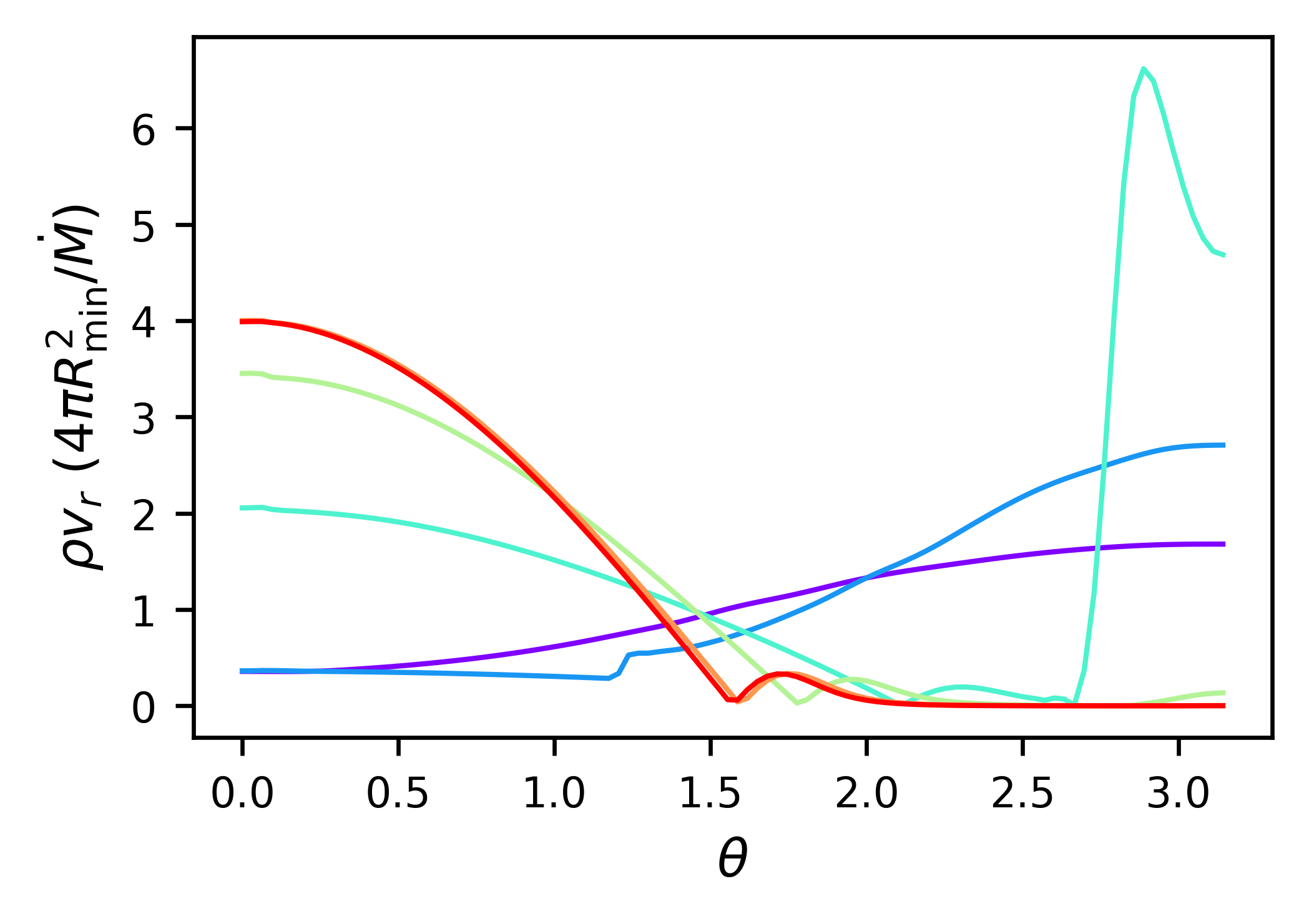}
                \parbox[b]{0.05em}{(e) \\[4ex] } \\
        \end{minipage}
        \hspace{0.05\textwidth} 
        \begin{minipage}[t]{0.45\textwidth}
                \centering
        \textbf{$\mathcal{M}_\infty = 1.5$}
                \includegraphics[width=0.85\textwidth]{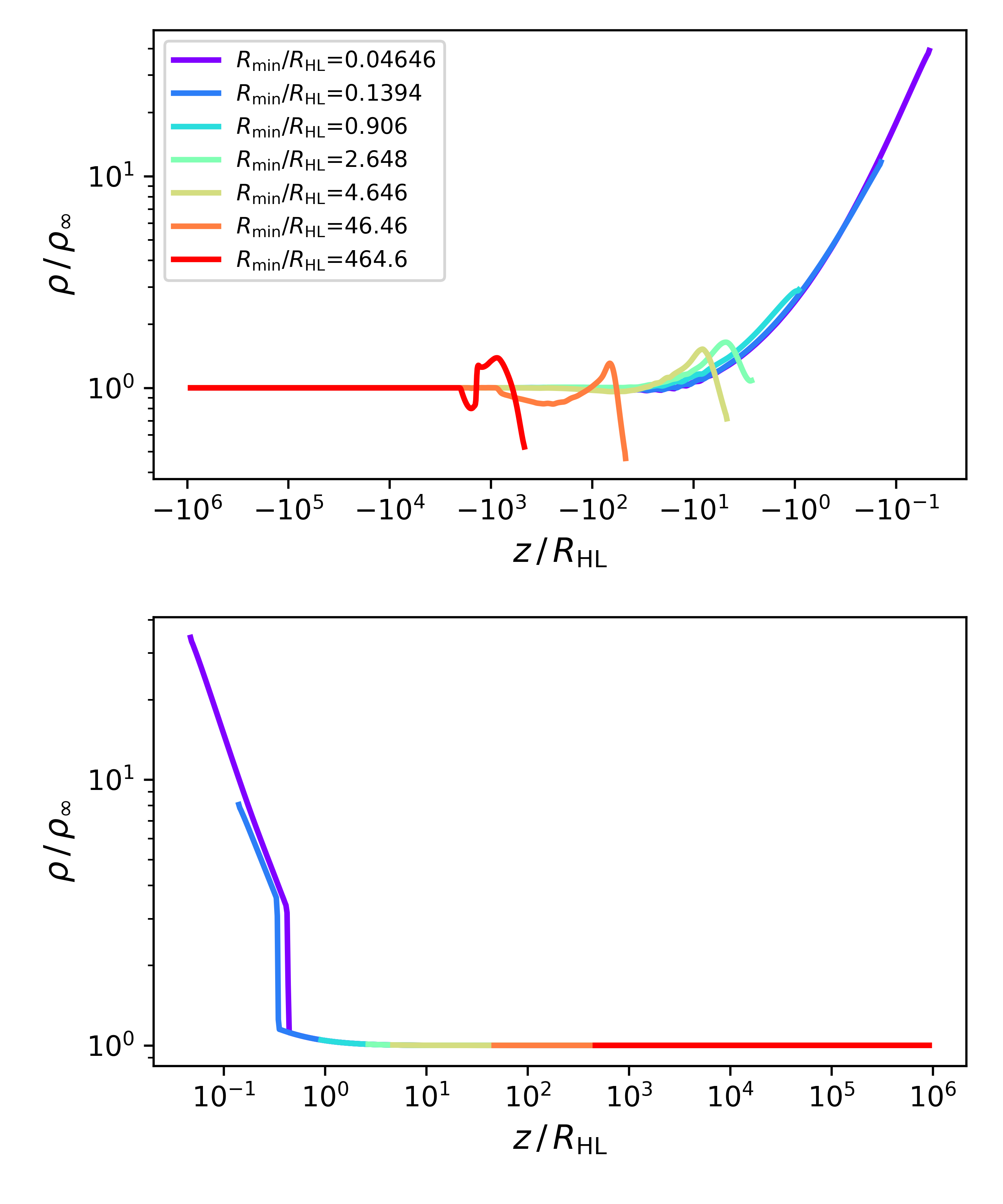}
                \parbox[b]{0.05em}{(f) \\[24ex] (g) \\[3ex]} \\
                \vspace{0.1cm}
                \includegraphics[width=0.85\textwidth]{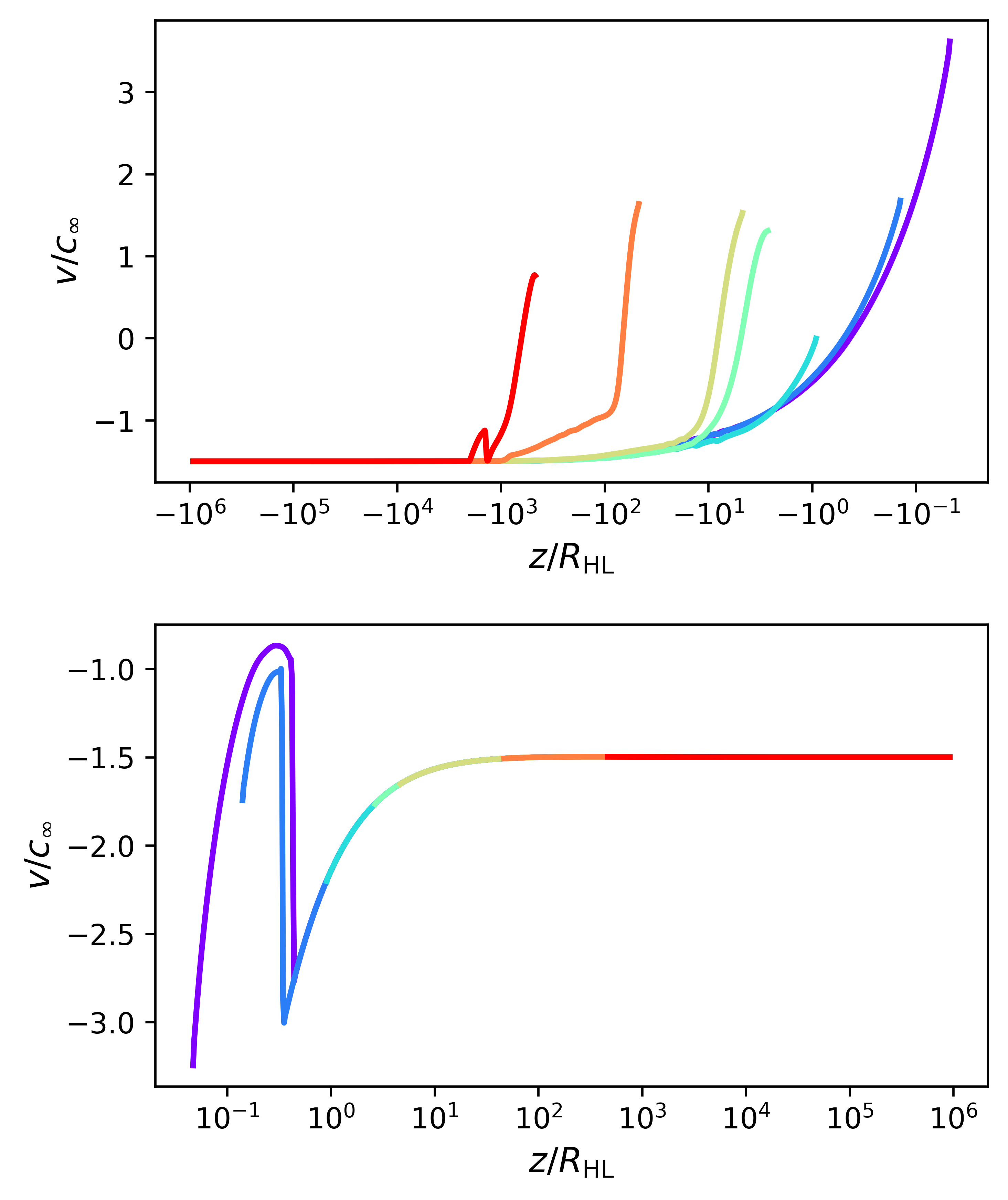}
                \parbox[b]{0.05em}{(h) \\[25ex] (i) \\[2ex]} \\
                \vspace{0.1cm}
                \includegraphics[width=0.85\textwidth]{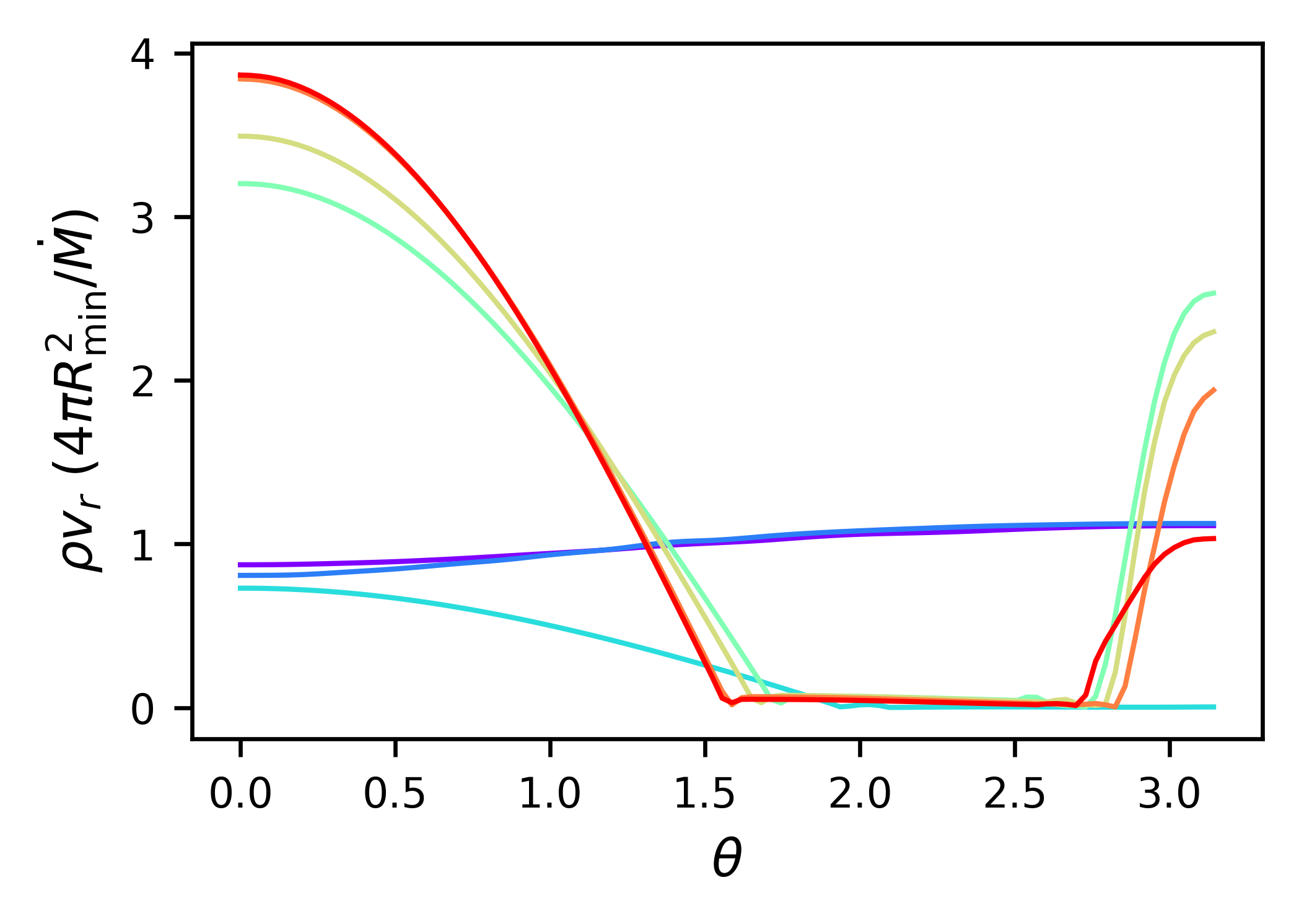}
                \parbox[b]{0.05em}{(j) \\[4ex] } \\
        \end{minipage}
        
        \caption{
        Flow analysis along the symmetry axis for $\mathcal{M}_\infty=10$ (left column) and $\mathcal{M}_\infty=1.5$ (right column). Density is normalized by the ambient density in (a), (b), (f), and (g), and velocity is normalized by the sound speed of the incoming gas in (c), (d), (h), and (i), both as a function of the normalized $z$-position.  The normalized, local accretion rate at $r = R_\mathrm{min}$ is shown as a function of the polar angle $\theta$ in (e) and (j).
        }
        \label{Analysis}
\end{figure*}

While we use $\mathcal{M}_\infty=10$ in Sect.~\ref{Result10} to clearly separate the length scales $R_{\mathrm{HL}}$ and $R_{\mathrm{Bondi}}$, we here present results for $\mathcal{M}_\infty=1.5$ to check for the impact of the Mach number and especially check the case where the Bondi radius and the Hoyle-Lyttleton radius are much closer to each other.
Results are shown in Fig.~\ref{Analysis}(d-f).

When comparing this scenario with $\mathcal{M}_\infty=10$, we basically obtain similar results. For both cases, the local accretion rate reaches a smooth profile for sufficiently small $R_\mathrm{min}$. 
The difference is observed in the accretion anisotropy. 
For $R_{\mathrm{min}}$ greater than $R_{\mathrm{HL}}$, for $\mathcal{M}_\infty=10$, there is zero accretion behind the accretor, in contrast to $\mathcal{M}_\infty=1.5$, where accretion still persists in this region; from the monotonic behaviour we can expect that this turns into zero accretion from behind (the geometrical limit) for even larger $R_{\mathrm{min}}$ though.
More interesting, for $R_{\mathrm{min}}$ smaller than $R_{\mathrm{HL}}$, there is a physical difference of the anisotropy between the Mach numbers. At $\mathcal{M}_\infty=1.5$, the larger stand-off distance results in a subsonic atmosphere, leading to nearly spherically symmetric accretion. With increasing Mach number, the kinetic effects increase and the anisotropy of the accretion increases accordingly.
The existence of a spherically symmetric atmosphere was also found by \citet{2016A&A...589A..10T} for supersonic non-accreting objects and by \citet{10.1093/mnras/stad3405} for subsonic accretors. 

\begin{figure*}[htbp]
        \centering
        \begin{minipage}[t]{0.45\textwidth}
                \centering
                \includegraphics[height=0.7\textwidth]{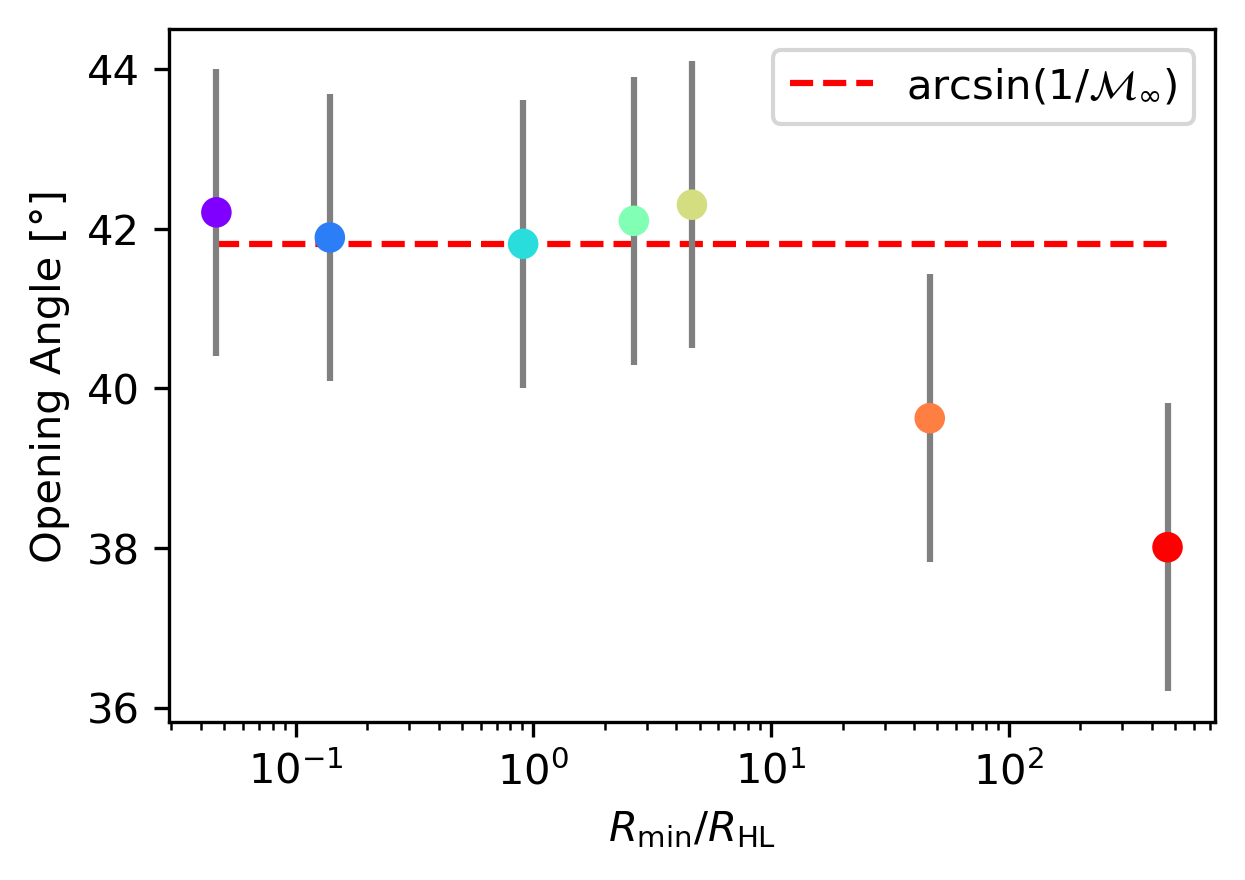}
                (a) \\ 
    \hfill
        \end{minipage}\qquad
        \begin{minipage}[t]{0.45\textwidth}
                \centering
                \includegraphics[height=0.7\textwidth]{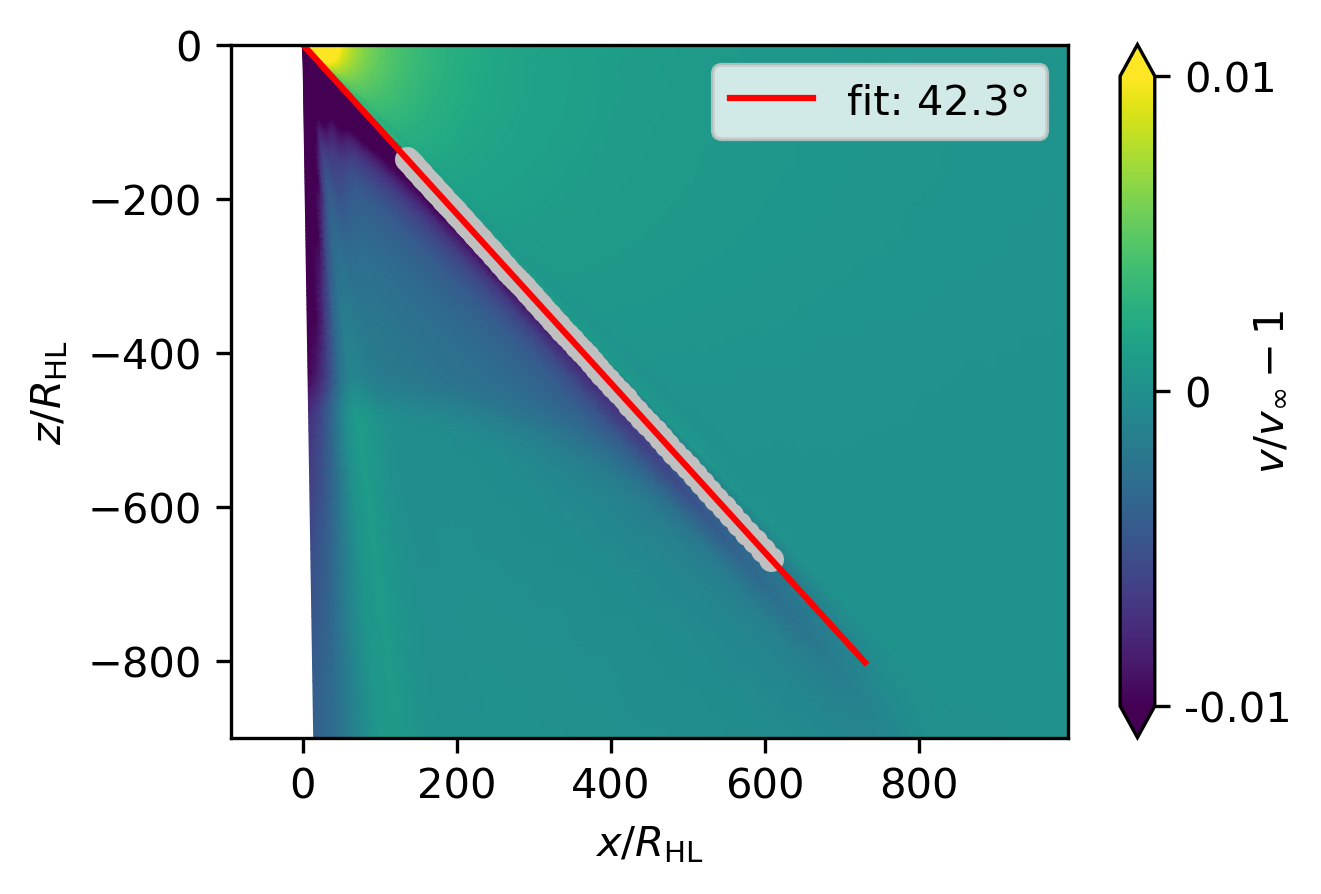}
                (b) \\ 
        \end{minipage}
        \caption{Mach cone analysis. Panel (a): For $\mathcal M_\infty=1.5$, the opening angle of the Mach-like cone is close to the expected value for not too large $R_\text{min}$. The error bars display the lattice resolution. Panel (b): Example ($R_\text{min}=4.646\, R_\text{HL}$) of determining the angle by fitting a straight line to a part of the shock front (grey points) far from the accretor.}
\label{opening-angle}
\end{figure*}

Figure \ref{opening-angle} displays the opening angle of the Mach-like cone for $\mathcal M_\infty=1.5$ and various $R_\mathrm{min}$. For each simulation, we applied a regression line to points on the straight part of the shock front (jump in $v$). 
For $\mathcal{M}_\infty=1.5$, the angles match well with $\arcsin(1/\mathcal{M_\infty})$, with deviations around 1\% for small $R_\mathrm{min}$. These deviations are much smaller than those reported in previous studies: 25\% for Mach 3 in \citet{1994ApJ...427..351R} and a factor of 2 in \citet{1989ApJ...336..313P}. For higher $R_\mathrm{min}$ the deviations are 5.2\% and 9\%. 
For $\mathcal M_\infty = 10 $, the values obtained were too high by almost 40\%, but they should be treated with caution anyway because they are less than $5\Delta\theta$.



\subsection{Accretion rates}
\label{Accretion rate}
\begin{figure}[htbp]
    \centering
    \includegraphics[width=0.5\textwidth]{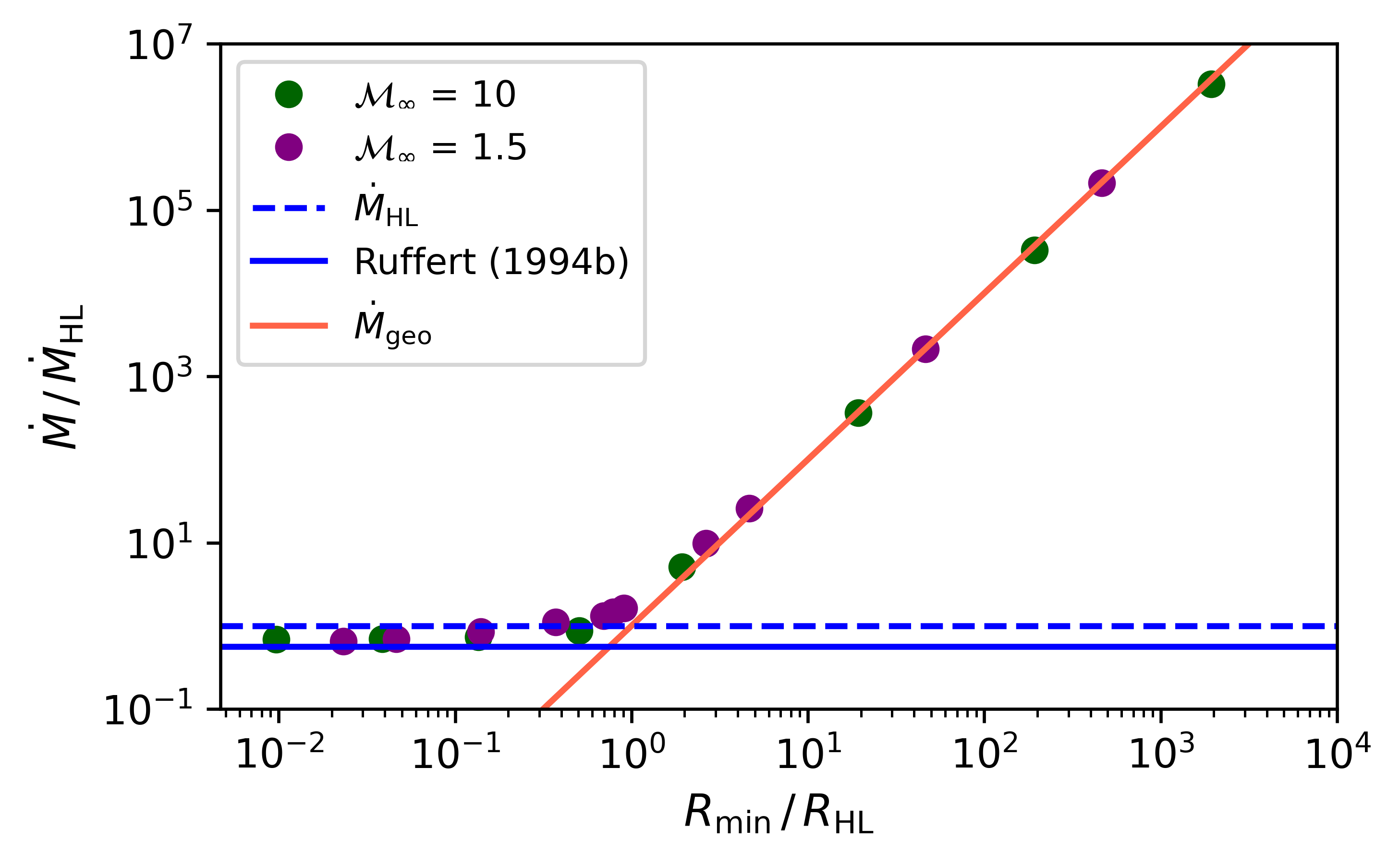}
    \caption{Mass accretion rate as a function of accretor size. Both axes are normalized by the \cite{1939PCPS...35..405H} accretion rate and radius. The dashed horizontal line at $\dot{M}/\dot{M}_{\mathrm{HL}} = 1$  represents the Hoyle-Lyttleton accretion rate, derived from Eq.~\eqref{HL-accretion}, the solid blue line the \citet{1994A&AS..106..505R} result $ \dot{M}/\dot{M}_{\mathrm{HL}}=0.446$, according to their Eq.~(12) for $ \gamma = 5/3 \Rightarrow \beta = 5/9$, and the red line the geometrical accretion rate according to Eqs.~\eqref{geo-accretion} and \eqref{norm-eq}. The results are presented for simulations conducted with two different Mach numbers, denoted as $\mathcal{M}_\infty$. }
    \label{fig:accretion}
\end{figure}

Figure \ref{fig:accretion} illustrates the variation of the measured accretion rate $\dot M$, normalized by $\dot{M}_{\mathrm{HL}}$ from Eq.~\eqref{HL-accretion}, against $R_{\mathrm{min}}$ normalized by $R_{\mathrm{HL}}$ from Eq.~\eqref{eq:defRdyn} for two simulation series with Mach numbers of $\mathcal{M}_\infty = 10$ and $1.5$, respectively.

Two limiting behaviours are evident, with the turnover point occurring at $R_{\mathrm{min}} \approx R_{\mathrm{HL}}$:
For $R_{\mathrm{min}}$ smaller than $R_{\mathrm{HL}}$, the accretion rate closely approximates $\dot{M}_{\mathrm{HL}}$, the \citet{1939PCPS...35..405H} accretion shown by the dashed line. In terms of convergence behaviour, the resulting flow is becoming independent of the value of $R_{\mathrm{min}}$ for $R_{\mathrm{min}} \ll R_{\mathrm{HL}}$.
Hence, we refer to the simulations with the smallest accretor's size as `fully resolved simulations' from here on. 
In our case, which includes pressure, it might be surprising that for the smallest $R_{\mathrm{min}}$, the Bondi-Hoyle model does not provide a better match than the Hoyle-Lyttleton model. However, it is important to note that both models are approximations.
As shown in Fig.~\ref{fig:accretion}, we included Ruffert's result \citep{1994A&AS..106..505R}  with $\beta(\gamma = 5/3) = 5/9$; however, this does not yield a perfect match either.

On the other hand, $\dot M$ approaches $\dot{M}_{\mathrm{geo}}$ for $R_{\mathrm{min}} \gg R_{\mathrm{HL}}$, because the gas particles hit the accretor before gravity becomes significant. Since everything but the radii cancels out in the normalized accretion rate, in this case
\begin{equation}
  \frac{\dot{M}}{\dot{M}_{\mathrm{HL}}} \approx \frac{\dot{M}_{\mathrm{geo}}}{\dot{M}_{\mathrm{HL}}}  =  \left (\frac{R_{\mathrm{min}}}{R_{\mathrm{HL}}} \right)^2 \,,
  \label{norm-eq}
\end{equation}
the data for both Mach numbers collapse onto this power-law line (shown in red in Fig.~\ref{fig:accretion}).

\section{Summary}
\label{summary}
A runaway accretor in supersonic, non-relativistic motion causes the formation of a bow shock and a Mach cone, even in the case of an accretor without a solid surface. The numerical size of the accretor changes the morphology of these structures; the larger the size, the more suppressed the relevant physical effects around the object are and thus the less they are resolved. 

Comparing the length scales with the accretor radius, if the accretor is smaller than the stand-off distance, a bow shock forms in front of the object, and the region between the stand-off distance and the accretor is filled with a slightly asymmetric `atmosphere'; the level of asymmetry depends on the Mach number of the flow, among other factors. 
Although the stand-off distance determines the bow shock position, the non-resolution of this structure has no impact on the flow morphology at larger scales. If we consider an accretor with a smaller radius than the Hoyle-Lyttleton radius, we can observe a Mach cone. For radii larger than the Hoyle-Lyttleton radius, only a Mach-like cone behind the object is observed. For the physics considered in the study, no higher-density trail is found at $\theta=\pi$;  a higher-density trail is only found beyond the Mach cone. 

The steady-state accretion rate depends on the accretor size as well and exhibits two regimes. One scenario arises when the accretor is smaller than the Hoyle-Lyttleton radius, and in this case the accretion rates closely align with the Hoyle-Lyttleton rates given by \citet{1939PCPS...35..405H}. Conversely, when the accretor's radius exceeds the Hoyle-Lyttleton radius, the accretion rates approach the geometrical limit.

In fully resolved simulations where the accretor's size is smaller than the stand-off distance, the level of anisotropy in the accretion rate follows the expectation of a slightly higher accretion from behind the object. 
In simulations with an accretor size larger than the Hoyle-Lyttleton radius, a much stronger deviation from spherical symmetry is evident, with accretion only from the incoming direction, as expected in the geometrical limit. Here, a `shadow' of low density forms behind the object.

\begin{acknowledgements}
    We thank Shu-ichiro Inutsuka for fruitful discussions about the modelling aspects of runaway black holes. 
    We thank Dennis Wehner for his contributions in analysing the Mach cone. 
    We thank the referee for their valuable comments and suggestions that improved the manuscript.
    R.K.~acknowledges financial support via the Heisenberg Research Grant funded by the Deutsche Forschungsgemeinschaft (DFG, German Research Foundation) under grant no.~KU 2849/9, project no.~445783058.
    R.K.~also acknowledges financial support from the JSPS Invitational Fellowship for Research in Japan under the Fellowship ID S20156; the idea for this study was conceived during a poster session at the EANAM9 conference, which R.K.~attended as part of the fellowship.
\end{acknowledgements}

\bibliographystyle{aa}
\bibliography{run_away.bbl}

\begin{appendix} 

\section{Higher mass for the accretor}
\label{sect:mass-scaling}
Because the mass of the accretor only enters the physical problem on hand in the definition of the escape speed and hence the Hoyle-Lyttleton radius, we expect the resulting flow morphology to be independent of the value of the mass when the results are normalized by this length scale. 
To check this expectation explicitly, and as an easy test of the simulation code and analysis scripts, we performed simulations for two masses of the object, $M = 1 M_\odot$ (as presented above) and additionally $M = 10^4M_\odot$.

\begin{figure}[htbp]
    \centering
    \includegraphics[width=0.5\textwidth]{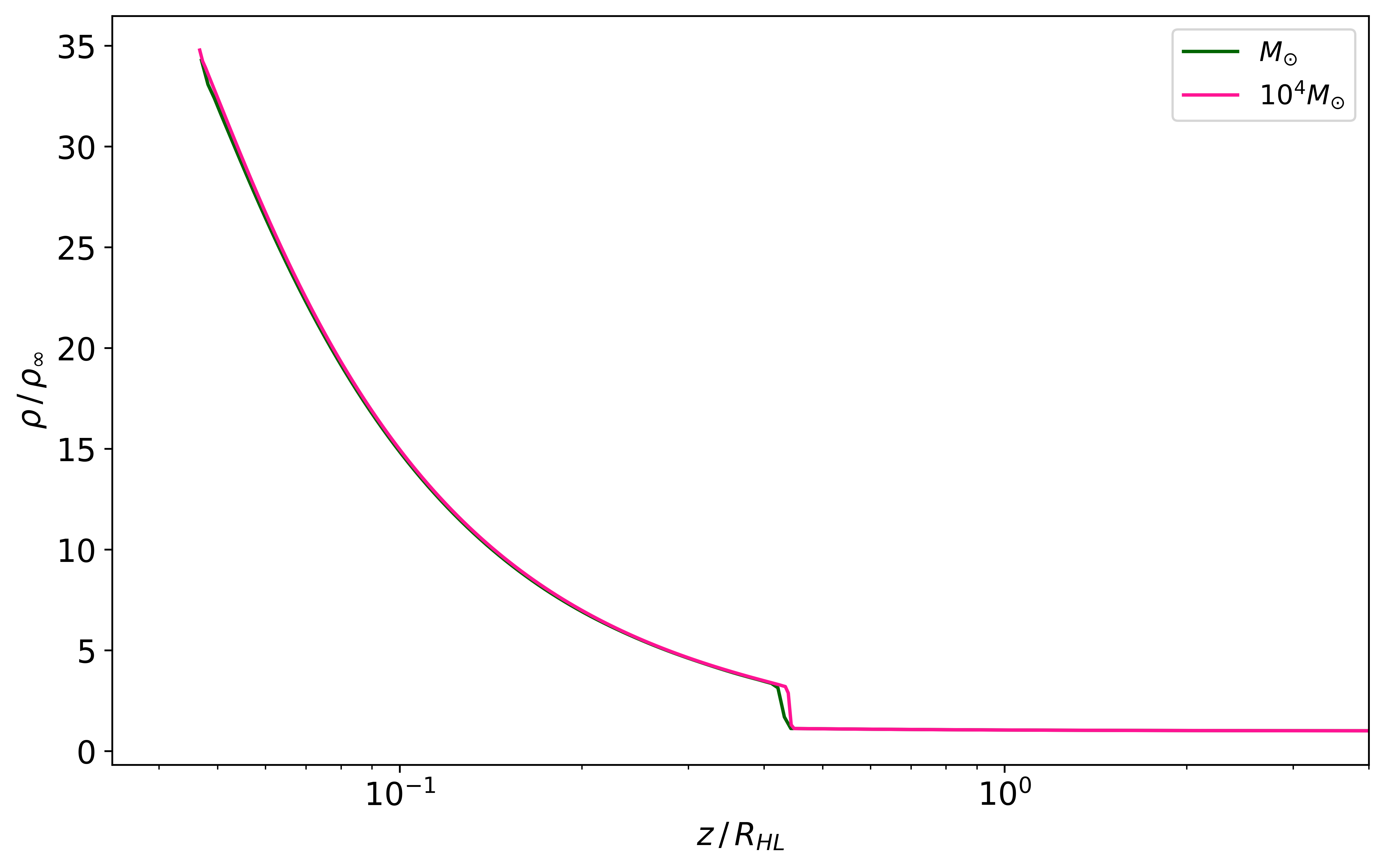}
    \caption{Normalized density distribution for an angle $\theta = 0$, representing the region in front of the object after the system reaches a steady state, depicted for two different masses. The Mach number is $1.5$.}
    \label{Mass}
\end{figure}

As illustrated in Fig.~\ref{Mass}, the results depict the density distribution in front of the accretor to check for the resulting distance of the bow shock and the density profile within this stand-off distance. 
The two simulation grids were chosen to be identical, and both are in units of $R_\mathrm{HL}$. 

The residuals of density and pressure between the two simulations were also calculated. The maximum residual for density was 1\%, and for pressure, it was 2.5\%, except at the shock front, which can be seen in Fig.~\ref{Mass}.

As anticipated, the results demonstrate that the object's mass does not impact the formation of a bow shock or any other flow morphology. This observation is consistent with the predictions of the Euler equations, suggesting that the object's mass serves as a scaling parameter and does not directly influence the characteristics of the flow.

\end{appendix}
\end{document}